\documentstyle[a4,12pt]{article}
\begin{document}
\begin{titlepage}
\begin{center}
\hspace*{10cm} TUM-HEP-261/96\\
\hspace*{10cm} LMU-TPW 96-31\\
\hspace{10cm} hep-th/9611060\\
\hspace{10cm} October 1996\\
\vspace*{1cm}

{\LARGE T-duality for open strings with respect to 
non-abelian isometries\footnote{Talk given by 
S.\ F\"orste at the ``Workshop on Gauge Theories,   
Applied Supersymmetry and Quantum Gravity'', 
Imperial College, London, UK, 5 - 10 July 1996}}\\

 \vspace{1.2cm}

{\large Stefan F\"orste$^{\S}$
\footnote{ E-Mail:Stefan.Foerste@physik.uni-muenchen.de}, 
Alexandros A.\ Kehagias$^{\sharp}$
\footnote{E-Mail: kehagias@physik.tu-muenchen.de}
and Stefan Schwager$^{\S}$
 \footnote{E-Mail:Stefan.Schwager@physik.uni-muenchen.de}}\\
\vspace{.8cm}

${}^{\S}$
{ Sektion Physik\\
Universit\"at M\"unchen\\
Theresienstra\ss e 37, 80333 M\"unchen\\
Germany}\\
\vspace{.6cm}

${}^\sharp${ Physik Department \\
Technische Universit\"at M\"unchen\\
D-85748 Garching, Germany}\\
\end{center}
\vspace{1cm}

\begin{center}
{\large Abstract}
\end{center}
We gauge the non-abelian isometries of a sigma model with boundaries. 
Forcing the field strength of the gauge fields to vanish renders the gauged 
model equivalent to the ungauged
one provided that boundary conditions are taken into account properly. 
Integrating out
the gauge fields gives the T-dual model. We observe that T-duality 
interchanges
Neumann (or mixed) boundary conditions with Dirichlet boundary conditions.
\end{titlepage}
\newpage
\section{Introduction}
Some time ago it has been observed that T-duality in toroidally compactified
open strings 
interchanges Neumann with Dirichlet boundary conditions \cite{polch}. In the
recent past  
there  has been renewed interest in this subject since it provides extended
objects 
(D-branes) which are important in the context of string dualities, (for an
excellent review see 
\cite{notes}). Therefore it is of special interest to generalize the T-duality
transformation 
for open strings living in a more general background.

The first step in this direction has been done in \cite{klimcik} where
backgrounds possessing 
a Poisson-Lie symmetry have been considered.  In \cite{alv}  and 
independently in 
\cite{do} T-duality was carried out in general backgrounds with Abelian
isometries.  
The present talk is addressed to the generalization for backgrounds with
non-Abelian 
isometries, work published in \cite{us} will be reported. 

Some papers dealing with T-duality corresponding to non-Abelian 
isometries are listed  
in \cite{quev,alot}. Here we will follow the initial work of
\cite{quev}, and we restrict ourself to semi simple isometry
groups. 

In the second section some preliminary consideration discussing boundary
conditions 
in a setting suitable for general backgrounds will be presented. 
How to gauge a
non-Abelian isometry in a sigma model with boundary  will be shown in the
third section. Finally, the T-duality will be performed in the fourth section,
followed by 
concluding remarks in a fifth section.
\section{Preliminary consideration} \label{sec:prem}
In this section we discuss how boundary conditions transform under 
T-duality.
For simplicity we will consider just one free boson on a world sheet with
boundary, 
the generalization to less trivial cases will be straightforward,
\begin{equation}
S = \int_{\Sigma} d^2z\, \partial_a X \partial^a X.
\end{equation}
This action is invariant under constant shifts in $X$. The first step in
deriving the T-dual 
model is to gauge this global symmetry, (for a review on T-duality see
\cite{grp}). In order 
to obtain an action invariant under the local transformation
\begin{equation}
X \rightarrow X + f(z) 
\end{equation}
we have to introduce two dimensional gauge fields $\Omega_a$ transforming
according to 
\begin{equation}
\Omega_a \rightarrow \Omega_a - \partial_a f .
\end{equation}
and to replace partial derivatives by covariant ones,
\begin{equation}
\partial_a X \rightarrow D_a X = \partial_a X + \Omega_a .
\end{equation}
A gauge invariant action is given by, (we neglect global issues since these 
can not 
be discussed in the non-Abelian case \cite{alot}),
\begin{equation}
S_{gauged} = \int_{\Sigma} d^2z\, \left( D_a X D^a X + \lambda F\right) +
\oint_{\partial \Sigma} 
ds\, c t^a \Omega_a,
\end{equation}
where $F$ is the field strength corresponding to the isometry gauge 
field $\Omega$, 
\begin{equation}
F = \epsilon^{ab}\partial_a \Omega_b,
\end{equation}  $\lambda$ is a Lagrange 
multiplier forcing the field strength $F$ to vanish, $t$ is the tangent vector
on the 
boundary $\partial \Sigma$ and $c$ is an arbitrary constant. Now, we consider
the 
partition function
\begin{equation}
Z = \int {\cal D}\Omega{\cal D}X {\cal D}\lambda e ^{-S_{gauged}}.
\end{equation}
Integrating out the Lagrange multiplier $\lambda$ forces the field strength $F$
to 
vanish which in turn constrains the gauge field $\Omega$ 
to be ``pure gauge'', i.e. $\Omega_a = \partial_a \rho$.
Upon the redefinition
\begin{equation} \label{shift}
X \rightarrow X + \rho
\end{equation}
the original model is obtained back. However, since we are 
considering a model
with boundary we have to be careful that the shift (\ref{shift}) does not
change the  
boundary condition. Therefore, let us specify the boundary condition in the
original 
model by
\begin{equation}
b^a\partial_a X_{| \partial \Sigma} = 0.
\end{equation}
(If the vector $b$ is normal to the boundary these are Neumann conditions 
whereas $b$ being tangent to the boundary gives - up to a constant -
Dirichlet conditions.) The (local) equivalence to the original model can  be
ensured 
by adding a second Lagrange multiplier term 
$\oint ds\, \kappa b^a\Omega_a$ 
on the boundary. Now, the gauged action (after partial integration) reads
\begin{equation} \label{gg} \begin{array}{l l}  
S_{gauged} = & \int d^2 z\, \left( \partial_a X\partial^a X +\Omega_a\Omega^a +2
\Omega_a 
\partial^a X - \epsilon^{ab}\partial_a\lambda\Omega_b\right)\\
& + \oint ds \left(c t_a + \kappa b_a + t_a \lambda\right) \Omega^a. 
\end{array} \end{equation}
The T-dual model is obtained by integrating out the gauge fields $\Omega$.
The action (\ref{gg}) is written such that it contains no derivatives 
of $\Omega$. 
Hence, the $\Omega$ integral is ultra local and  
integrations over gauge fields
with arguments in the bulk and over gauge fields with arguments on the 
boundary
factorize \cite{do},
\begin{equation}
\int {\cal D}\Omega_{\Sigma \, \cup \, \partial \Sigma}\left( \dots\right) =
\int {\cal D}\Omega_{\Sigma }\left( \dots\right)\times
\int {\cal D}\Omega_{ \partial \Sigma}\left( \dots\right).
\end{equation}
Integrating out gauge fields where the argument is in the bulk
of the world sheet gives the dual action,
$\tilde{S} = \int d^2z\, \partial_a \lambda \partial^a \lambda$.
Since on the boundary $\Omega$ appears linearly that integration gives
a two dimensional delta function
\begin{equation} \label{delta}
\delta^{(2)} \left( ct_a + \kappa b_a + \lambda t_a\right),
\end{equation}
specifying the boundary conditions of the dual model as we will see now.
There are essentially two different cases to be considered,
\newcounter{cases}
\begin{list}{\bf (\arabic{cases}) }{\usecounter{cases}}
\item $ b \, \not\!\|\, \,  t$
\item $b\; \, \parallel \, t$ .\label{case_b}
\end{list}
In case {\bf (1)} the arguments 
of the two-dimensional delta function
can vanish simultaneously only if $\kappa = 0$, and we are left with the
dual boundary condition
\begin{equation}  \label{dirich}
\lambda_{|\partial\Sigma} = - c.
\end{equation}
In case {\bf (2)} the two-dimensional delta function just 
identifies the
second Lagrange multiplier $\kappa$ with the negative of the first one
$\lambda$ on the 
boundary, (in this case we can put the constant $c$ to zero since the tangent
component of the gauge field is zero),
\begin{equation}
\kappa = -\lambda_{|\partial \Sigma}.
\end{equation}
So, the final picture we see is
described as follows:
\newcounter{faelle}
\begin{list}{\bf (\arabic{faelle})}{\usecounter{faelle}}
\item We start with Neumann or mixed boundary conditions. 
This can be
thought of as obtained from a general open string sigma model by 
requiring the
variation of the action to vanish for free varying ends of the string ($\delta
X_{|\partial \Sigma} = any$), (the target space metric
is required to be non degenerate). In the dual model we obtain 
Dirichlet boundary conditions
(\ref{dirich}) enforced by a delta function, i.e.\ $\delta
\lambda_{|\partial \Sigma} = 0$.    
\item We start with Dirichlet boundary conditions 
(up to a constant)
enforced by a delta function, i.e.\ the variations of the end of the string are
frozen to vanish. In the dual model no boundary conditions are specified. So,
a natural choice is to pick those boundary conditions 
obtained from an action
principle with free varying ends.
\end{list}
We have seen in this simple setting that 
an open string with free varying ends
is T-dual to an open string whose ends are fixed on a surface in the target
space, the D-brane. 
\section{Gauging the non-Abelian isometries}
The general sigma model describing the action of an open string in a non
trivial background is given by, (since any consistent open string theory
includes closed strings there are also a metric and an antisymmetric tensor
field),
\begin{equation}\label{openaction}
\begin{array}{ll}
S=&-\frac{1}{2}
\int_\Sigma d^2\sigma \left(G_{mn}\partial_a X^m\partial^aX^n +
B_{mn}\epsilon^{ab}\partial_aX^m\partial_bX^n\right) \\
&-\int_{\partial\Sigma} ds A_m\partial_sX^m
+\frac{1}{4\pi}\int_\Sigma d^2\sigma\, \sqrt{\gamma}
R^{(2)}\Phi -\frac{1}{2\pi}\int_{\partial\Sigma} ds k \Phi \, , 
\end{array}\end{equation}
where $\Phi $ is the dilaton coupling to the Gauss-Bonnet density of
the world sheet,
i.e.\ to the scalar curvature $R$ in the bulk and to the geodesic
curvature $k$ on the boundary.
We assume that there are isometries, i.e.\ a set of Killing vectors
${\cal L}_{\xi_I} G_{mn} = 0$
satisfying a semi simple Lie-algebra,
$\left[ \xi_I, \xi_J\right] = {f_{IJ}}^K \xi_K $.
Now we are going to specify the conditions under which
the sigma model (\ref{openaction}) is invariant with respect
to the global transformations
\begin{equation} \label{eq:global}
\delta X^m = \epsilon^I \xi^m_I ,
\end{equation} 
with $\epsilon^I$ being the constant transformation
parameters. The first condition is that the Lie-derivative of
the dilaton $\Phi$ has to vanish, and from now on
we are excluding the dilaton from the discussion and will
return to it in the conclusion when mentioning open problems.
The Lie-derivatives of the antisymmetric tensor $B$ and
the $U(1)$-gauge field $A$ have to vanish up to
gauge transformations, i.e.
\begin{equation} \label{eq:first-con}
\begin{array}{l l}
{\cal L}_{\xi_I} B_{mn} = &\partial_m\omega_{In} -
\partial_n\omega_{Im}, \\
{\cal L}_{\xi_I} A_m= & \partial_m\phi_I +\omega_{Im},
\end{array}
\end{equation}
where the $\omega$ and $\phi$ are arbitrary
functions. The rhs of (\ref{eq:first-con}) is invariant
under the redefinitions, ($ \omega_I = \omega_{Im}dX^m$),
\begin{equation} \label{eq:invar}
\begin{array}{ll}
\omega_{I}^\prime = & \omega_I + dh_I ,\\
\phi^{\prime}_I = & \phi_I - h_I + k_I ,
\end{array}
\end{equation}
where the $h_I$ are arbitrary functions and
the $k_I$ are constants. By evaluating the
Lie-derivatives with respect to a commutator of
two Killing vectors in two different ways we derive
the following consistency conditions,
\begin{equation}
\begin{array}{ll}
{\cal L}_{\xi_I}\omega_J-{\cal L}_{\xi_J}\omega_I = &
{f_{IJ}}^K\omega_K +d\rho_{IJ},\\
{\cal L}_{\xi_I}\phi_J - {\cal L}_{\xi_J}\phi_I = &
{f_{IJ}}^K\phi_K - \rho_{IJ},
\end{array}
\end{equation}
with $\rho$ being some function. From \cite{hull1, hull2}
we know that in the closed string case an
anomaly free gauging is possible only if $d\rho_{IJ}$
vanishes. So, we have to remove $\rho$ by employing
the symmetry transformations (\ref{eq:invar}). This can be
done provided $d\rho$ satisfies certain integrability
conditions \cite{hull1}. We assume that this is the case
and set $d\rho_{IJ}$ to zero, i.e.\ $\rho_{IJ} = k_{IJ}$
to a constant. The possible symmetry transformations 
we are left with are constant shifts in $\phi_I$ under
which the $k_{IJ}$ transform. For simplicity 
we take the antisymmetric tensor $B$ to be zero from
now on.

Now we are going to gauge the model, i.e.\  $\epsilon $
in (\ref{eq:global}) depends on the position on the world sheet.
We introduce isometry gauge fields $\Omega$ transforming
as a connection
\begin{equation}
\delta \Omega_{a}^J = \partial_a \epsilon^I +{f_{KJ}}^I\Omega_{a}^K
\epsilon^J .
\end{equation}
Since we have put the antisymmetric tensor to zero 
the bulk part is gauged just by replacing partial
derivatives with covariant ones,
\begin{equation}
D_a X^m = \partial_a X^m - \xi^{m}_I
\Omega^{I}_a .
\end{equation}
In order to gauge the boundary part we use Noether's method,
i.e.\ we add the contribution
\begin{equation}
S^{(1)} =\int ds C_I \Omega^{I}_s,
\end{equation} 
where the index $s$ denotes the tangent component. Requiring 
gauge invariance leads to two equations for $C_I$, {\it viz.}
\begin{equation} \label{eq:noether}
\begin{array}{ll}
C_I = & -A_m\xi^{m}_I +\phi_I +\lambda_I ,\\
{\cal L}_{\xi_I} C_J = & -{f_{JI}}^KC_K ,
\end{array} \end{equation} 
with $\lambda_I$ constant. The two equations
in (\ref{eq:noether}) are compatible provided that
\begin{equation}
k_{IJ} -{f_{IJ}}^K\lambda_K =0
\end{equation}
implying an integrability condition on the 
constants $k_{IJ}$,
\begin{equation}
{f_{IJ}}^K k_{KL} + cycl. \; permutations =0 .
\end{equation}
\section{Non-Abelian T-duality}
It is useful to split the coordinates into those which
transform non trivially under the isometry gauge group
carrying a Latin index and spectators carrying a Greek
index. Using light cone coordinates on the world sheet,
(for conventions cf \cite{us}), the gauged action reads
\begin{equation} \label{eq:gaga}\begin{array}{ll}
S\left( X, \Omega\right) = & \int_{\Sigma} d^2z \left\{
G_{\alpha\beta}\partial X^\alpha \bar{\partial}X^\beta +
G_{\alpha m}\left( \partial X^\alpha \bar{D} X^m +
\bar{\partial}X^\alpha D X^m\right) \right.\\
&\left. + G_{mn} DX^m\bar{D}X^n + \Lambda_I G^I\right\} \\
&^ - \int_{\partial \Sigma} ds \left( A_\alpha \partial_s X^\alpha +
A_m\partial_s X^m - C_I \Omega ^{I}_s \right) ,
\end{array}\end{equation}
where $G = \partial \bar{\Omega}-\bar{\partial{\Omega}} +[\Omega,
\bar{\Omega}]$ is the field strength of the isometry gauge fields and the
$C_I$ are determined according to the previous section. The gauged
action (\ref{eq:gaga}) is gauge invariant provided that the
Lagrange multiplier $\Lambda$ transforms in the adjoint, $\delta\Lambda_I 
= - {f_{IJ}}^K\Lambda_K\epsilon^J$.
Here, we start with free varying ends of the string, i.e.\
Neumann or mixed boundary conditions. As we have seen
in section \ref{sec:prem} a (in principle necessary) second 
boundary-Lagrange multiplier drops out finally, and therefore
we drop it already here. The first Lagrange multiplier forces
the gauge field excitations to be pure gauge (locally) and 
after absorbing the gauge parameters in a field redefinition
one obtains the original model back. The T-dual is constructed
by integrating out the gauge fields. Again, we rewrite the action
such that there are no derivatives of the isometry-gauge fields
and employ the ultra locality to factorize the integral into a 
product of an integral with field arguments in the bulk and
an integral with field arguments on the boundary. The bulk
integral will provide the dual action coinciding with the
closed string result \cite{quev}. The boundary integral
specifies Dirichlet conditions on the dual coordinates
enforced by a delta function,
\begin{equation}
\Lambda_I + C_I \, _{|\partial\Sigma} = 0
\end{equation}
being covariant with respect to the isometry transformations.
In addition one has to fix the gauge by fixing the $X^m$ and
in general also some of the $\Lambda$'s. This concludes
the construction of the T-dual model.
\section{Conclusions}
We have gauged the non-Abelian isometries in a 
general sigma model with boundary and constructed
the T-dual model by integrating out the gauge fields.
We observed that in the open string case T-duality
interchanges Neumann or mixed boundary conditions
(corresponding to free varying ends) with Dirichlet
conditions enforced by a delta function, i.e.\ the 
variations of the ends of the string are fixed in the
dual model.

Let us finish this talk with some open problems. First of
all there are global issues which are hard to address also
in the closed string case. In addition one should mention
the dilaton. In a general sigma model with boundary
there could be in principle two dilatons, one coupling to
the curvature in the bulk and one coupling to the
geodesic curvature on the boundary. A string
interpretation requires only one dilaton coupling 
to the Gauss-Bonnet density, (that this is consistent
with renormalizability has been shown in \cite{behrndt}).
Now, a natural guess for the T-dual model would be that
the dilaton shift is the same as in the closed string
case \cite{quev,arkady}, and that in the dual model the
dilaton couples just to the Gauss-Bonnet density
as well. A proof of that guess is missing so far.

After this talk has been given ref.\ [14] appeared where
results differing from ours have been obtained in a 
canonical transformation framework. The reason is that
in the canonical transformation scheme one takes frequently
variational derivatives and in ref.\ [14] periodic
boundary conditions are imposed on variations. 
The canonical transformation and the method presented here
are actually in agreement as discussed in \cite{buckow}.  
\section*{Acknowledgments}
S.\ F.\ is supported by GIF- German Israeli foundation for
scientific research, A.\ K.\ is supported by the
Alexander von Humboldt foundation
and S.\ S.\ is supported by DFG - Deutsche Forschungs
Gemeinschaft. This work is partly supported by the EC programs
SC1-CT92-0789 and the European Comission TMR programs
ERBFMRX-CT96-0045 and ERBFMRX-CT96-0090.


\begin{thebibliography}{15}
\bibitem{polch} J. Dai, R. G. Leigh and J. Polchinski, Mod. Phys. Lett. 
{\bf A4}(19989)2073;\\
P.\ Ho\v{r}ava, Nucl. Phys. {\bf B327}(1989)461, Phys. 
Lett. {\bf B231}(1989)251
\bibitem{notes} J.\ Polchinski , S. Chaudhuri 
and C. V. Johnson, {\it ``Notes on D-branes''}, hep-th/9602052
\bibitem{klimcik}C. Klim\v{c}\'{\i}k, P. Severa, Phys. Lett. {\bf B376}(1996)82;
Phys. Lett. {\bf B381}(1996)281; {\it ``Open strings and D-branes in
WZNW model''}, hep-th/9609112
\bibitem{alv}E. Alvarez, J.L.F. Barbon and  J. Borlaf, {\it ``T-duality for
open strings''}, Nucl. Phys. {\bf B}(in press), hep-th/9603089
\bibitem{do}H. Dorn and  H.-J. Otto, Phys. Lett. {\bf B381}(1996)81
\bibitem{us} S. F\"orste, A. A. Kehagias and S. Schwager, 
Nucl. Phys. {\bf B478}(1996)141, hep-th/9604013
\bibitem{quev} X. de l'Ossa and F. Quevedo, Nucl. Phys. {\bf B403}(1993)377
\bibitem{alot}E. Alvarez, L. Alvarez-Gaum\'{e} and Y. Lozano, Nucl. Phys.
{\bf B424}(1994)155;\\
A. Giveon and E. Kiritsis, Nucl. Phys. {\bf B411}(1994)487;\\
A. Giveon and M. Ro\v{c}ek, Nucl. Phys. {\bf B421}(1994)173;\\
S. Elitzur, A. Giveon, E. Rabinovici, A. Schwimmer, G. Veneziano, 
Nucl. Phys. {\bf 435}(1995)147 \\
N. Mohammedi, Phys. Lett. {\bf B375}(1996)149 
\bibitem{grp} A. Giveon, E. Rabinovici, M. Porrati, Rhys. Rep. {\bf 244}(1994)77
\bibitem{hull1} C. Hull and B. Spence,  Phys. Lett. {\bf B232}(1989)204
\bibitem{hull2}I. Jack, D. R. T. Jones, N. Mohammedi and H. Osborn,
Nucl. Phys. {\bf B332}(1990)332 
\bibitem{behrndt} K. Behrndt and H. Dorn, Int.J.Mod.Phys. {\bf A7}(1992)1375
\bibitem{arkady} A. A. Tseytlin, Mod. Phys. Lett. {\bf A6}(1991)1721
\bibitem{yolanda}J. Borlaf and Y. Lozano, {\it ``Aspects of T duality in open
strings''}, FTUAM-96-26, hep-th/9607051
\bibitem{buckow}S. F\"orste, A. A. Kehagias and S. Schwager, 
{\it ``Non Abelian T-duality for open strings''}, 
Nucl. Phys. (Proc. Suppl.){\bf B}
to appear, hep-th/9610062
\end{thebibliography}
\end{document}